# Transient Scattering from an Axisymmetrical Perfectly Conducting Object with a Dielectric Coating

I. G. Efimova[1] and Z. V. Sedel'nikova[2]

[1]Moscow State University
[2] International Research Institute of Management Sciences

November 14, 2003

*Abstract*---Scattering of the Gaussian video pulse by a perfectly conducting circular cone coated with dielectric is considered. The problem is solved using the method of integral equations in the low and resonance frequency bands followed by applying the Fourier transform to obtain time-domain responses. Backscattered fields are investigated for different permittivities of the dielectric in the case of the axial incidence of a plane wave.

Diffraction of nonstationary electromagnetic fields can be studied using either the time- or frequency-domain approach [1, 2]. A proper choice of the technique depends on the purpose of investigations. For example, when different bodies are successively illuminated by one and the same pulse, the time-domain methods are most suitable for analyzing the influence of the shape of a scatterer on the response. However, when various incident pulses are scattered by the same body, it is more efficient to employ the frequency-domain methods. In this case, the frequency-domain scattering response (FSR) is calculated only once, and, then, the Fourier transform is applied to the product of the FSR and the spectrum of an incident signal, which appears to be a good instrument to analyze the influence of the shape of the incident pulse on scattered fields.

In this paper, we numerically investigate nonstationary wave diffraction by a perfectly conducting circular cone with a homogeneous dielectric coating (permittivity $\varepsilon$ is constant over the dielectric layer) in a homogeneous isotropic medium using the method of frequency-domain integral equations [3]. Consider the case when a video pulse illuminating a cone with $2\alpha = 23^\circ$, $r = 0.32a$, and $d = 0.6a$ ($2\alpha$ is the angle at the vertex, $a$ is the radius of the metal base, $r$ is the radius of the vertex and base edge roundings, and $d$ is the thickness of the dielectric layer), propagates along its axis towards the vertex.

Back-scattered field $E(\omega)$ can be represented as the product of the amplitude and phase frequency responses denoted by $A(\omega)$ and $\varphi(\omega)$, respectively,

$$E(\omega) = A(\omega) \exp[i\,\varphi(\omega)]. \qquad (1)$$

Let the incident signal be a smooth video pulse, for example, of the Gaussian form

$$f(t, z) = \exp[g^2(z - t)], \qquad (2)$$

where $t$ is time and $g$ is the parameter which specifies pulse duration $\tau$ at the level $f(t, z) = 1/e$, i.e., $\tau = 2/g$. Calculating the product of the incident signal spectrum and the frequency function of the back-scattered field and applying the Fourier transform to the result, we obtain the scattered field as a function of time

$$\mathring{A}^s(t) = 1/(2\pi) \int_{-\infty}^{\infty} A(\omega)\, S_f(\omega) \cos[\omega t + \varphi(\omega)]\, d\omega, \qquad (3)$$

where $S_f(\omega)$ is the spectrum of the incident signal. The spectrum of the Gaussian pulse has the form

$$S_f(\omega) = (\sqrt{p}/g) \exp[-\omega^2/(4g^2)]. \qquad (4)$$

Since the numerical algorithm is developed for solving the system of frequency-domain integral equations for a coated body in a finite frequency band, one has to replace the infinite limits of integral in (3) by finite ones, which is equivalent to truncating the spectrum of the incident signal. This limitation does not result in considerable errors of computations because, in this paper, we consider a Gaussian pulse with a rather narrow spectrum. When necessary, the neglected part of the incident signal spectrum can be estimated using the quantity $\omega a/c$ ($c$ is the speed of light in free space) involved in calculations of $E(\omega)$. In the case under consideration, $\omega_{max} a/c = 2.25$ and

$$\int_{\omega_{max}}^{\infty} S_f(\omega)d\omega \Big/ \int_{0}^{\infty} S_f(\omega)\,d\omega = 0{,}00152. \qquad (5)$$

In what follows, we consider a Gaussian pulse with $\tau = 4а/ñ$.

The figure shows transient responses (backscattered fields) of coated and perfectly conducting cones calculated in the case when the Gaussian pulse propagates along the axis of symmetry towards the vertex.

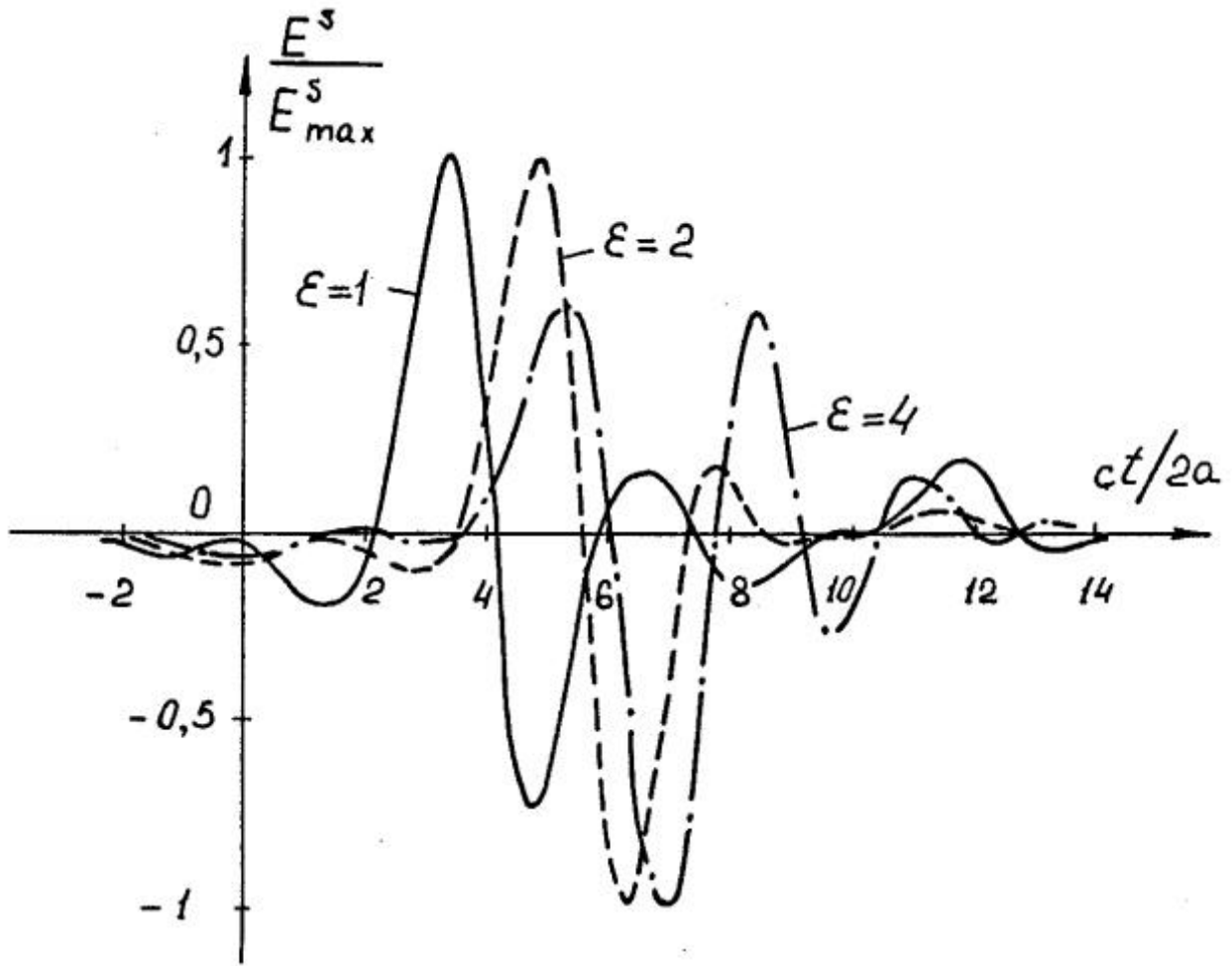

The zero moment is fixed so that the incident signal achieves its maximum at the origin when $t = 0$; i.e., $f(0, 0) = 1$. Similarly to scattering of a rectangular video pulse or a radio pulse by a perfectly conducting cone [2], in the case of a coated cone, one can also separate fields due to the first- and second-order diffraction from the edge of the metal base. The pulse primarily diffracted from the edge has an extremum at the moment $ct/2a \cong 3.4$, corresponding to the interval during which the wave covers the forward and return paths from the origin to the plane where the curvature of the body has a jump discontinuity between the lateral surface and the rounded edge of the cone. Note that the circle containing this curvature discontinuity is one of the scattering centers of the cone.

The pulse diffracted by the edge for the second time is characterized by a smaller amplitude, the opposite sign, and the time delay $\tilde{n}\Delta t/2a \cong 1.4$ with respect to the pulse considered above. This interval is longer than time $\Delta t_{12}$ necessary for the creeping wave to cover the distance between two opposite points on the shadow base surface ($\tilde{n}\Delta t_{12}/2a = 1$), because, when the incident pulse is rather long ($c\tau/a = 4$), the pulses due to the first- and second-order diffraction overlap.

The presence of a dielectric layer results in certain changes in the backscattered field as compared to the field scattered by a perfectly conducting cone. At $\varepsilon = 2$, the pulse due to the first-order diffraction has an extremum at the moment $ct/2a = 4.9$. Note that the signal propagates at speed $c$ in free space, while the path in the dielectric layer is covered at speed **v** ($c/\varepsilon^{1/2} \leq v < c$). The amplitude of the second-order diffraction pulse increases and, at $\varepsilon = 2$, becomes equal to the amplitude of the pulse primarily diffracted from the edge. This effect can be explained as follows. In the presence of a dielectric layer, a considerable portion of the energy of the primarily diffracted field is transformed into the energy of a quasi-surface wave, which propagates in the layer coating the base and, then, is again scattered by the edge. This portion of the energy increases with permittivity, and, at $\varepsilon = 4$, the field intensity corresponding to the second-order diffraction considerably exceeds the intensity of the primarily diffracted signal. As $\varepsilon$ increases, the second-order diffraction field is transformed into a quasi-surface wave, which propagates in the layer on the base and, then, is again scattered by the edge. Thus, the third-order diffraction occurs. When $\varepsilon$ is sufficiently large, the process is repeated, and one can observe pulses with decreasing amplitudes, which correspond to the diffraction of subsequent orders. For example, in the response of the cone covered by a layer with $\varepsilon = 4$ (the dashed-and-dotted curve in the figure), in addition to the third-order pulse with a considerable amplitude, we can see two decreasing pulses due to the diffraction of the fourth and fifth orders.

The reflection by the lateral surface and rounded vertex of the cone also contributes to the response. However, the intensity of the signal due to these factors is much smaller than the intensity of the field scattered by the edge of the cone base. As $\varepsilon$ grows, the field reflected by the lateral surface and vertex noticeably decreases.